\documentclass[%
 aps,
 prx,%
 amsmath,amssymb,floatfix,superscriptaddress,
 reprint,%
]{revtex4-2}
\usepackage{nameref}
\usepackage{soul}
\usepackage{hyperref}
\usepackage{graphicx}% Include figure files
\usepackage{bm}% bold math
\usepackage{color}
\usepackage[bf]{subfigure}
\usepackage[mathlines]{lineno}% Enable numbering of text and display math
\begin{document}

\title{Signatures of brain criticality unveiled by maximum entropy analysis across cortical states}

\author{Nastaran Lotfi}

\email{nastaran@df.ufpe.br}
\affiliation{Departamento de Física, Universidade Federal de Pernambuco, Recife, PE 50670-901, Brazil}
\author{Antonio J. Fontenele}
\author{Tha{\'\i}s Feliciano}
\author{Leandro A. A. Aguiar}
\affiliation{Departamento de Física, Universidade Federal de Pernambuco, Recife, PE 50670-901, Brazil}

\author{Nivaldo A. P. de Vasconcelos}
\author{Carina Soares-Cunha}
\author{B\'arbara Coimbra}
\author{Ana Jo\~ao Rodrigues}
\author{Nuno Sousa}
\affiliation{Life and Health Sciences Research Institute (ICVS), School of Medicine, University of Minho, Braga, 4710-057, Portugal}
\affiliation{ICVS/3B’s - PT Government Associate Laboratory, Braga/Guimar\~aes, Portugal}
\author{Mauro Copelli}
\affiliation{Departamento de Física, Universidade Federal de Pernambuco, Recife, PE 50670-901, Brazil}
\author{Pedro V. Carelli}
\email{pedro.carelli@ufpe.br}
\affiliation{Departamento de Física, Universidade Federal de Pernambuco, Recife, PE 50670-901, Brazil}

\date{\today}
\begin{abstract}

It has recently been reported that statistical signatures of brain criticality, obtained from distributions of neuronal avalanches, can depend on the cortical state. 
We revisit these claims with a completely different and independent approach, employing a maximum entropy model to test whether signatures of criticality appear in urethane-anesthetized rats. 
To account for the spontaneous variation of cortical state, we parse the time series and perform the maximum entropy analysis as a function of the variability of the population spiking activity.
To compare data sets with different number of neurons, we define a normalized distance to criticality that takes into account the peak and width of the specific heat curve. 
We found an universal collapse of the normalized distance to criticality dependence on the cortical state on an animal by animal basis.
This indicates a universal dynamics and a critical point at an intermediate value of spiking variability.

\end{abstract}
%\begin{document}

\maketitle

\newpage

%\section{Introduction}

Since Beggs and Plenz first reported neuronal avalanches in cortical slices~\cite{beggs2003neuronal}, the critical brain hypothesis has gained support in experimental data and become an important paradigm to understand brain dynamics~\cite{beggs2007criticality,chialvo2010emergent,shew2013functional,munoz2018}. 
According to this hypothesis, the computational advantages of a brain poised at or near a second order phase transition are optimal transmission capacity~\cite{shew2011information}, largest repertoire~\cite{haldeman2005,shew2013functional} and maximum dynamic range~\cite{kinouchi2006optimal,shew2009neuronal,larremore2011predicting}, among others.

In their seminal work, Beggs and Plens have shown that the distribution of avalanche sizes in cultured slices of rat brain followed a power law with exponent 3/2. 
This exponent coincides with the one found for a critical branching process (or any other model in the mean-field directed percolation universality class)~\cite{beggs2003neuronal,shew2009neuronal}.
This was just one of several scale-invariant phenomena expected to occur at a critical point.

In the years that followed, however, the investigation of neuronal avalanches in less reduced preparations raised some controversy. 
On one hand, power-law avalanche size distributions of spiking activity could be easily found in vivo during synchronized states (characterized by slow LFP - Local Field Potential - oscillations) under ketamine-xylazine~\cite{ribeiro2010spike} and isoflurane~\cite{hahn2017spontaneous} anesthesia.
On the other hand, long-range time correlations could be observed only during desynchronized states (characterized by fast LFP oscillations) in freely behaving rats, but not under ketamine-xylazine anesthesia~\cite{ribeiro2010spike}.

Fontenele et al. have proposed a solution for the controversies between different data sets, by probing criticality across different cortical states~\cite{fontenele2019criticality}. 
It is well known that the degree of synchronizations in the brain varies with the behavioral state. 
In slow wave sleep, for instance, the cortical LFP activity has lower frequency and high amplitude, which corresponds to a synchronized state with high spiking variability. 
In an awake and attentive animal, the cortical LFP has high frequency and low amplitude ~\cite{steriade2013brainstem}, corresponding to a desynchronized state with low spiking variability. 
Fontenele \textit{et al.} have identified consistent markers of a phase transition at an intermediate level of spiking variability, where both avalanches and time correlations consistently satisfy more stringent scaling relations~\cite{fontenele2019criticality,friedman2012,touboul2017power}.

Here we investigate whether a similarly spike-variability-dependent analysis of neuronal data would reveal signatures of criticality under a completely different approach. 
We focus on maximum entropy models~\cite{jaynes1957information}, which consist in a methodology of extracting the desired statistics from limited data with a minimum of plausible assumptions. 
Bialek and collaborators have shown that maximum entropy models are an effective and parsimonious way of reconstructing higher-order statistics in neuronal dynamics, based on single-neuron firing rates and pairwise correlations~\cite{schneidman2006weak}. 
Later, other works have proposed that signatures of criticality could be unveiled in retinal data using the divergence of a generalized specific heat of the maximum entropy model built from the data~\cite{tkavcik2015thermodynamics,broderick2007faster,fries2015rhythms,bastos2015visual}.

Specifically, we use a maximum entropy model which is based on the firing rate of the network in different time steps~\cite{mora2015dynamical} to study criticality across cortical states in urethane-anesthetized rats. 
As done previously, here a cortical state will be characterized by a proxy, namely the coefficient of variation (CV) of the population firing rate~\cite{clement2008cyclic,harris2011cortical,de2017coupled,fontenele2019criticality}.
We divide the time series according to CV values and apply the maximum entropy method for each division, analyzing the family of specific heat curves as the urethanized brain drifts from more synchronous to less synchronous states.

%\section{Methods} \label{Sec:method}

%\subsection{Data acquisition}

The data used in this analysis are taken from two experimental setups, as described previously~\cite{fontenele2019criticality}. 
Spikes have been recorded from 32-(64-)channel silicon probes implanted in the primary visual cortex (V1) of urethane-anesthetized Long-Evans (Wistar-Han) rats up to a duration of 3~hours (see details in~\cite{SM}). %
%\subsection{Maximum Entropy analysis}
%\label{sec:MaximumEntropy}
From the data we extracted the binary spiking matrix $\{\sigma_{i,t}\}$ as follows: 
we divided the time series into windows of length $\Delta t$ ($20-50$~ms). 
If neuron $i$ has spiked at least once in a time window $t$, then $\sigma_{i,t}=1$, (otherwise, it is zero). 
In the Supplemental Material, we show that the results are robust with respect to the value of $\Delta t$ (Fig.~S3~\cite{SM}).

Since we want to address the differences in dynamical regimes observed in different cortical states, it is natural to employ a variant of the maximum entropy formalism that takes into account the dynamical nature of the spike trains. 
Following Mora \textit{et al.}~\cite{mora2015dynamical}, a Boltzmann-like distribution is defined,

\begin{equation}
 P_\beta ( \{ \sigma_{i,t} \} ) = \frac{1}{Z(\beta)} \exp (- \beta E(\{ \sigma_{i,t} \} ))\; , \label{eq:probability}
\end{equation}
where $Z(\beta)$ is the normalization constant, $E$ is the ``energy'' of the spike trains and $\beta\equiv 1/T$, a control parameter, is equivalent to an inverse temperature $T$ and it is set to 1 to describe the observed spike statistics. 
The idea of the method is to maximize the entropy $-\sum_{\{\sigma_{i,t}\}} P\log P$ subject to observable constraints in the data~\cite{jaynes1957information,schneidman2006weak}.

Being interested only in global phenomena and not in individual interaction between neurons, Mora \textit{et al.} proposed a maximum entropy model where the energy function depends only on the population firing rates and transitions between consecutive firing rates~\cite{mora2015dynamical}. 
In this way, the joint probability distributions of $K_t \equiv \sum_{i=1}^{N}\sigma_{i,t}$ at two different times $P_u(K_t,K_{t+v})$ are constrained, and the energy function is defined as:
\begin{equation}
 E=-\sum_{t}^L h(K_t) - \sum_{t}^L\sum_{u=1}^{v} J_u(K_t,K_{t+u})\; , \label{eq:E}
\end{equation}
where $N$ is the number of neurons, $L$ is the number of time bins and $v\geq 1$ is the temporal range of model. 
$h(K)$ and $J_u(K,K^\prime)$ are parameters which are fitted to the data using the technique of transfer matrix. 
We refer the reader to Appendices H and I of the Supplemental Material of Ref.~\cite{mora2015dynamical} for details.

Once $P_\beta$ is determined, the specific heat can be calculated as a function of $\beta$: 

\begin{equation}
 c({\beta})=\frac{\beta^2}{NL} \langle \delta E^2 \rangle_{\beta}\; , \label{eq:s_heat}
\end{equation}
where $\delta E \equiv E - \langle E \rangle_\beta$ is the fluctuation from the mean energy and its average is taken under $P_{\beta}$~\cite{mora2015dynamical}. 
Note that $P_\beta$ maximizes the entropy given the data only for $\beta=1$. 
By allowing $T$ to vary, a family of probability distributions is traversed, and a peak of $c$ that tends towards $T=1$ as $N$ increases is interpreted as a signature that the system is critical~\cite{tkavcik2015thermodynamics,mora2015dynamical}. 

%Since different datasets have different number of neurons, 
In order to handle the dependence of $c$ on the system size, %
%To show the dependence of $c$ on the system size, %
for each data set $N$ neurons were selected randomly from the total set of neurons recorded. 
This was repeated over 20 random selection of units and the shown result is their average. 
To control for the significance of the results, we also repeat the specific heat calculation for surrogate data, in which for each neuron the sequence of interspike intervals was shuffled (Supplemental Material Section B~\cite{SM}).

%\subsection{CV parsing}
%\label{CVparsing}

To understand how the spiking variability could affect the maximum entropy analysis, the data is segmented in windows of duration $W=10$~s.
For each window $i$, the coefficient of variation (\textit{CV}) of the population firing rate $K_t$ is calculated: 

\begin{equation}
    CV_i=\frac{\sigma_i}{\mu_i} \label{eq:cv},
\end{equation}
where $\mu_i$ is the mean and $\sigma$ is the standard deviation of $K_t$ within window $i$. 
To have better statistics for the model fitting, we concatenate 50 windows of similar \textit{CV}s, calculate their average $\langle CV \rangle$ and run the maximum entropy algorithm to find the heat capacity as a function of the temperature for different values of $\langle CV \rangle$.
Robustness of the results against changes in $W$ was also verified (Fig.~S5~\cite{SM}).

%\section{Results}
%\subsection{Analysis of the time series as a whole}

\begin{figure}[!ht]

    \includegraphics[width=0.95\linewidth]{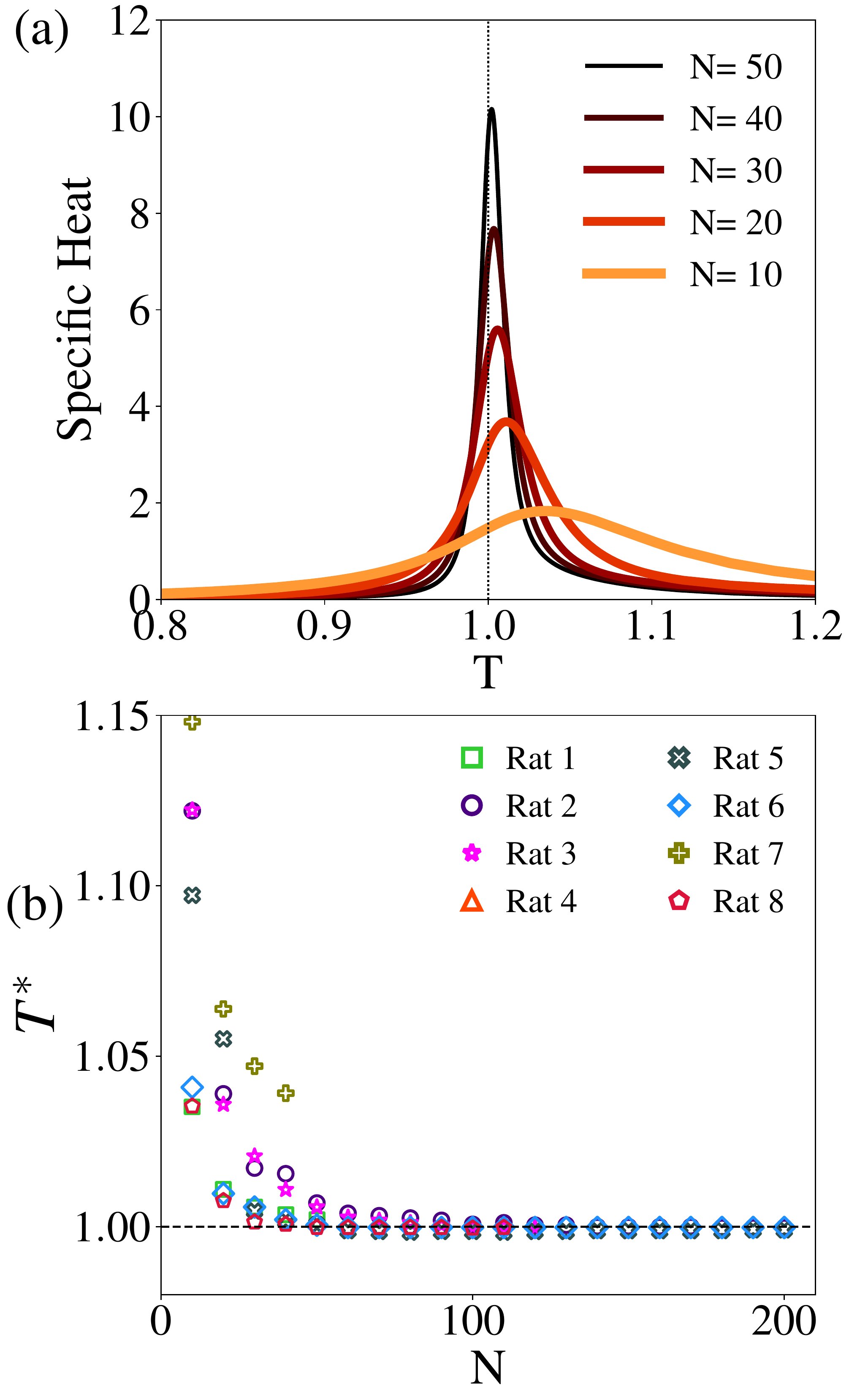}
    \caption{(a) Specific heat versus temperature ($T$) for an increasing number of neurons (Rat 1, which had 52 neurons recorded), with $\Delta t=20$~ms and $v=2$. 
     (b) variation of critical temperature ($T^*$) as a function of the number of neurons (N) considered to fit the model. 
    }
    \label{fig:rat_diff_n}
\end{figure}

We start our investigation by simply considering the whole time series of the data sets. 
Fitting the Maximum Entropy model to the data, we obtained curves of specific heat $c(T)$. 
$T^*$ is defined as the temperature at which $c$ is maximal. 
As we exemplify in Fig.~\ref{fig:rat_diff_n}a for a single rat, the larger the number $N$ of neurons, 
the closer $T^*$ was to $T=1$ and the larger the value of $c(T^*)$, suggesting a critical dynamics~\cite{tkavcik2015thermodynamics,mora2015dynamical}. 
These results are consistent across rats, as shown in Fig.~\ref{fig:rat_diff_n}b. %
When we repeat the analysis for surrogate (shuffled) data (see Supplemental Material Section B~\cite{SM}), the specific heat values are much smaller and the peaks occur for temperatures $T^*_{surrogate} < 1$ (Fig.~S1a). 
This suggests that surrogate data at $T=1$ is above the critical temperature, therefore in a disordered phase (as expected).

%\subsection{Analysis by spiking variability}

\begin{figure}[!ht]
    \includegraphics[width=0.95\linewidth]{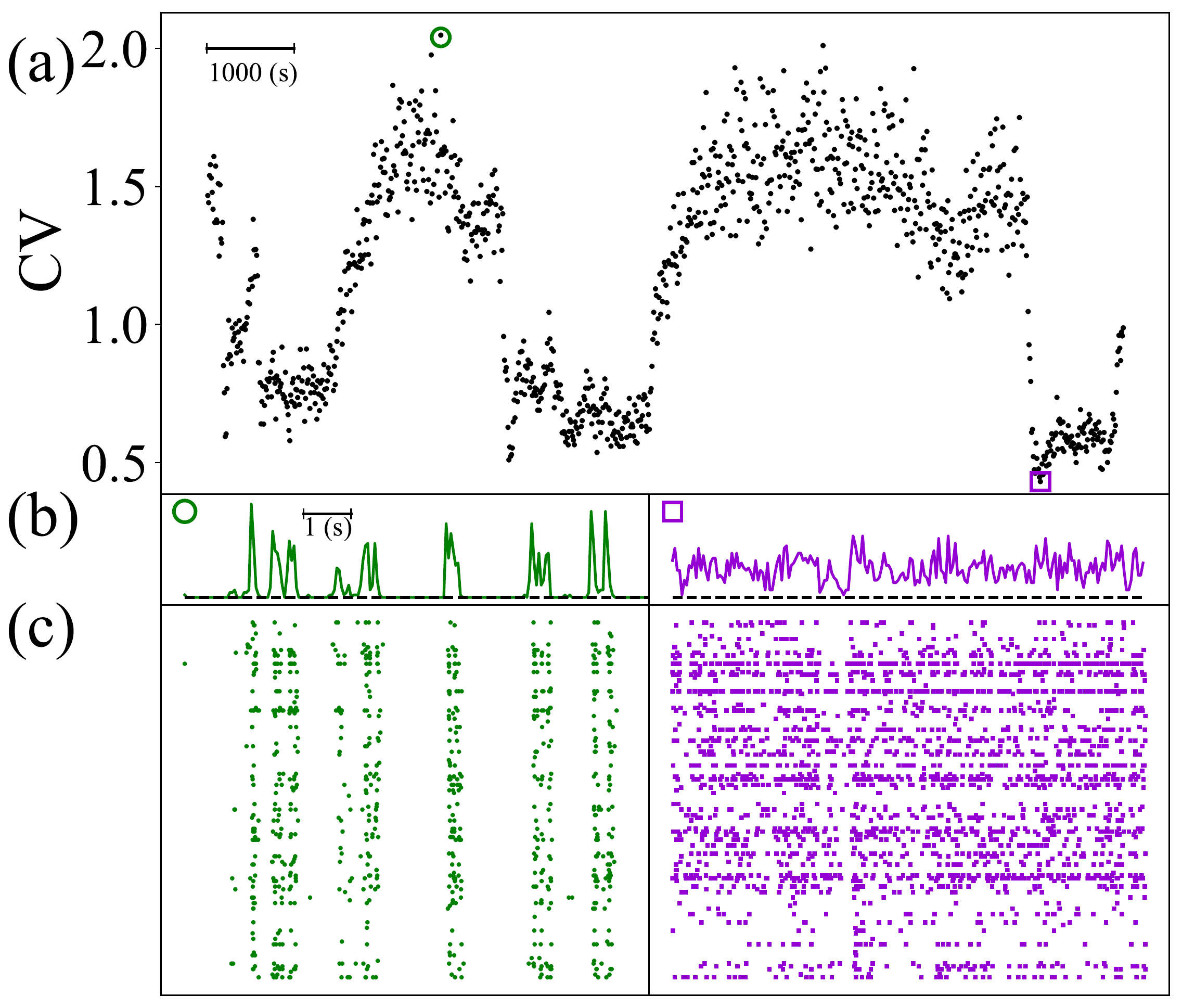}
    \caption{(a) \textit{CV} as a function of time for the whole time series (Rat 3) ($\Delta t=50$~ms, $W=10$~s). 
    The maximum and minimum values are highlighted by a circle and a square, respectively. 
    (b) Firing rate $K_t$ corresponding to the maximum (left) and minimum (right) CV values of the time series (dashed line is an indicator of zero). (c) Raster plots corresponding to (b), where each line represents a different neuron and each dot is a spike. Larger (smaller) values of CV correspond to more (less) synchronized states.}
    \label{fig:cv_3}
\end{figure}

These results, however, should be taken with a grain of salt. %(should be interpreted with caution)
Note that the model proposed by Mora \textit{et al.} assumes that the data is stationary~\cite{mora2015dynamical}, since the parameters $h$ and $J_u$ in the energy function of Eq.~\eqref{eq:E} are time-independent. 
But the dynamics of the spiking data in urethanized brains, on the other hand, changes considerably in the time scale of the whole record ($\sim 2-3$ hours). 
A common index to quantify these changes is the coefficient of variation ($CV$) of the population firing rate, which we calculate within windows of duration $W$. %$W=10$~s (see section~\ref{CVparsing}).
Fig.~\ref{fig:cv_3}a shows the time evolution of $CV$ for a single rat in the scale of hours, where one observes instances of very high spiking variability in more synchronized states ($CV \simeq 2$, left plots of Fig.~\ref{fig:cv_3}b~and~\ref{fig:cv_3}c), very low spiking variability in more desynchronized states ($CV \simeq 0.5$, right plots of Fig.~\ref{fig:cv_3}b~and~\ref{fig:cv_3}c), and pretty much everything in between. 
As shown in Fig.~S2~\cite{SM}, each experiment has its own, apparently unpredictable, evolution of $CV(t)$.

This lack of stationarity in longer time scales suggests that we are mixing together very different dynamical regimes when the Maximum Entropy analysis is applied to the whole time series. 
To reconcile the assumed hypothesis of stationarity of the model and the changes in cortical state in a slow (${\cal O}(>10~\mbox{s})$) time scale, we consider this analysis by previously segmenting and aggregating data by $CV$ values, in line with Fontenele \textit{et al.}~\cite{fontenele2019criticality}.

\begin{figure}[!tb]
    \centering
    \includegraphics[width=0.95\linewidth]{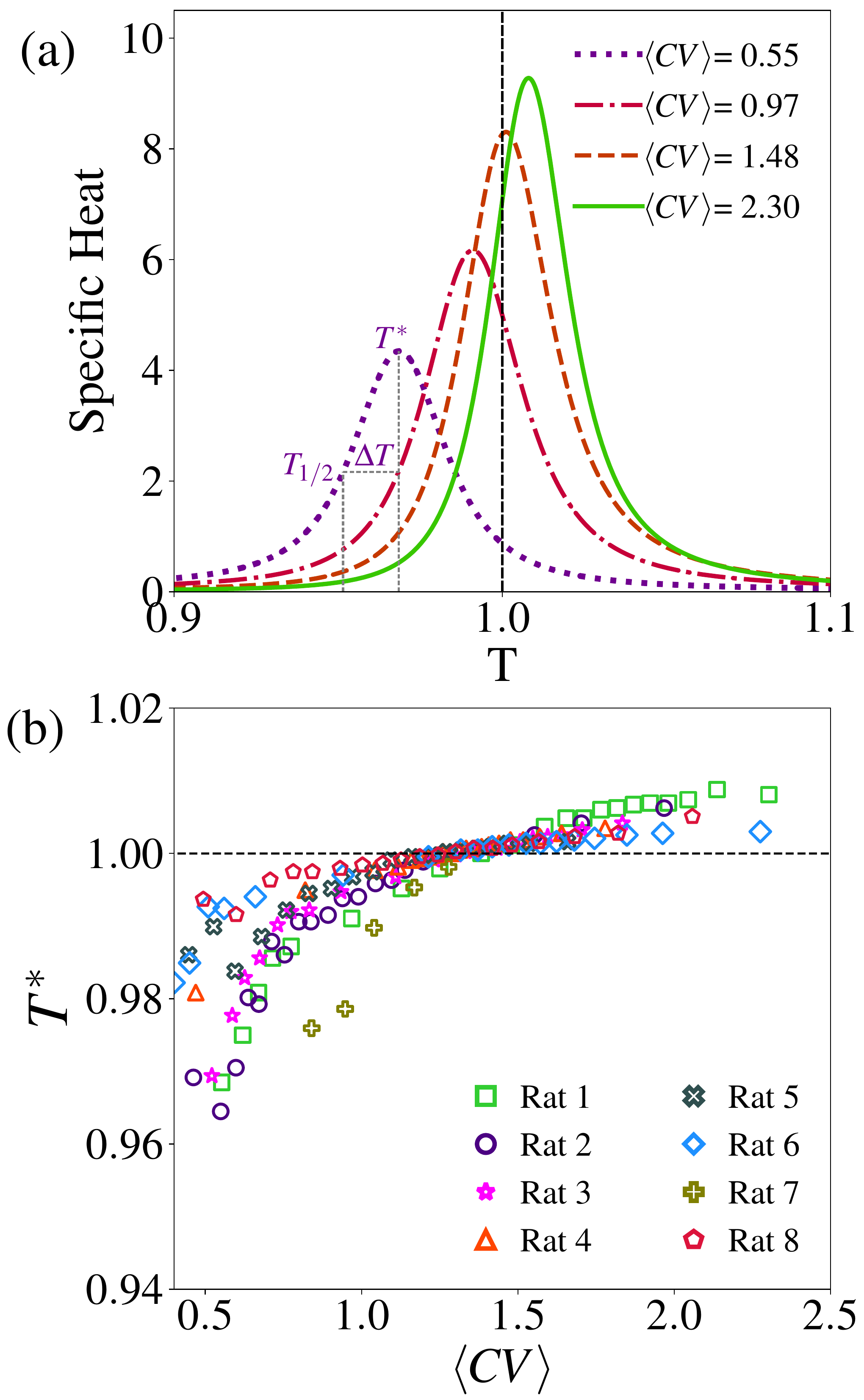}
    \caption{(a) Specific heat as a function of the temperature for different values of $\langle CV\rangle$ of Rat~1, $N=52$, $\Delta t=50$~ms). 
    The definitions of $\Delta T$ (Eq.~\eqref{eq:DeltaT}), $T^*$ and $T_{1/2}$ are illustrated for $\langle CV\rangle=0.55$ (see text for details). 
    (b) $T^*$ versus $\langle CV\rangle$ for different rats, where we have employed the maximal number of neurons in each data set.}
    \label{fig:rat_diff_cv}
\end{figure}

To do so, the maximum number of neurons in each data set is considered. 
In Fig.~\ref{fig:rat_diff_cv}a, the dependence of $c(T)$ with $\langle CV\rangle$ is shown for one data set (Rat 1; results for surrogate data are shown in Fig.~S1b~\cite{SM}). 
As the value of $\langle CV\rangle$ increases, the peak temperatures $T^*$ now increase from below $T=1$ to above it.

The interpretation of these results can be tricky. 
Note that the data is, by definition, described by the model at $T=1$. 
Whether $T=1$ is considered ``high'' or ``low'', i.e. whether the data corresponds respectively to a disordered or an ordered phase, depends on where the critical temperature $T^*$ lies. 
For low $\langle CV\rangle$ (desynchronized states), $T=1$ is higher than $T^*$, suggesting a disordered phase.
Accordingly, the high-$\langle CV\rangle$ (synchronized states) would correspond to the ordered phase.

\begin{figure}[t!ht]
    \centering
    \includegraphics[width=0.95\linewidth]{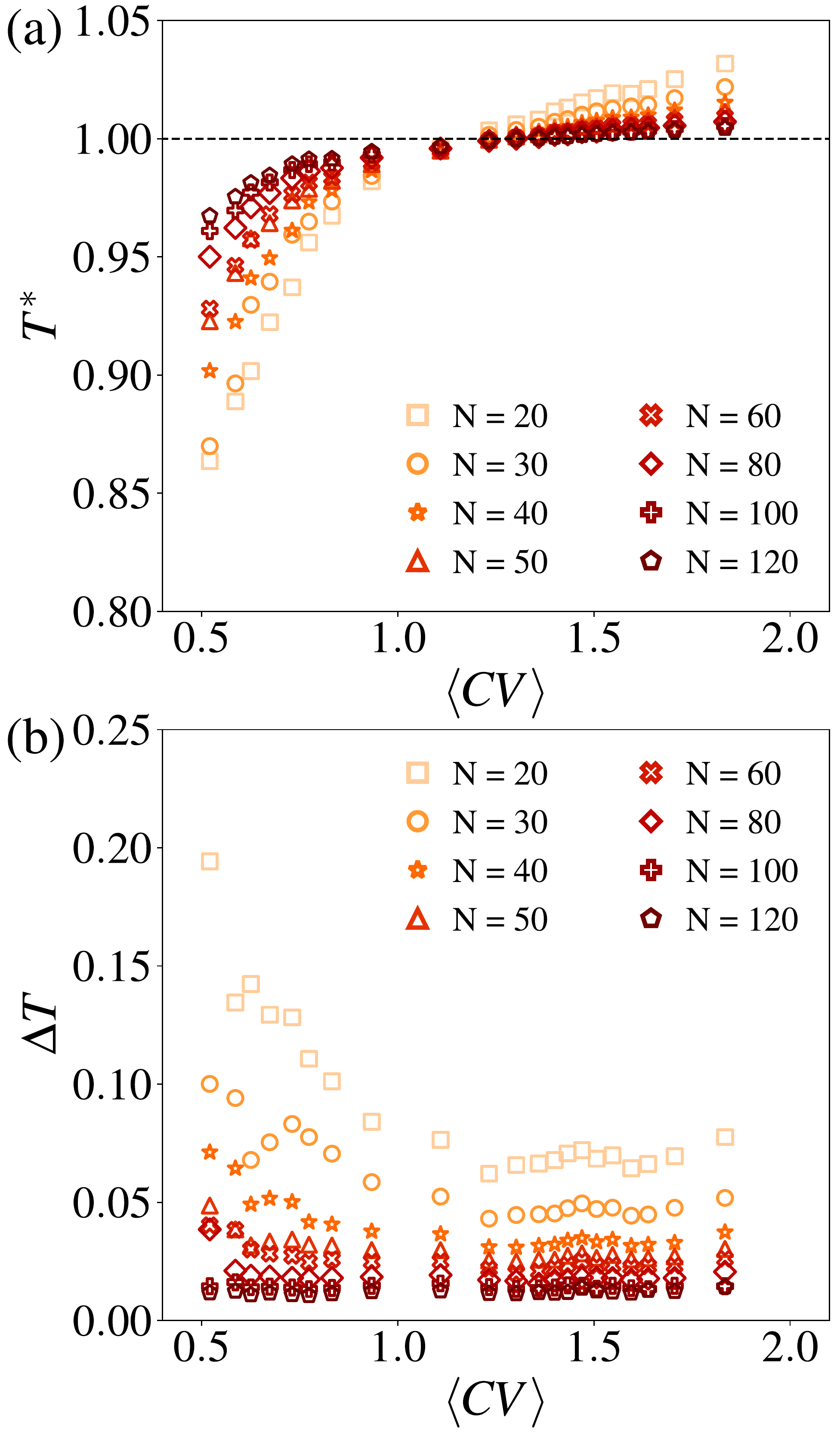}
    \caption{(a) Peak temperature $T^*$ versus $\langle CV\rangle$ for different number of neurons (Rat~3, $\Delta t=50$~ms). Note that $T^*$ approaches $T=1$ from below (above) for low (high) $\langle CV\rangle$. (b) Width $\Delta T$ (see Eq.~\ref{eq:DeltaT} and Fig.~\ref{fig:rat_diff_cv}a) of the specific heat curve versus $\langle CV\rangle$ for different number of neurons.} 
    \label{fig:rat_diff_nn}
\end{figure}

Therefore, if we parse the data by spiking variability, the signatures of criticality do not appear in the whole time series.
As shown in Fig.~\ref{fig:rat_diff_cv}b, the peak temperature $T^*$ coincides with $T=1$ only in a narrow range of $\langle CV\rangle$. 
These results are robust across animals, and suggest a critical point between the desynchronized and the synchronized extremes, as reported by Fontenele \textit{et al.}~\cite{fontenele2019criticality}.

Of course, different rats have different number of recorded neurons, and we would like to understand whether those differences can be controlled for when analyzing the $\langle CV\rangle$ dependence of $c(T)$ curves, such as those of Fig.~\ref{fig:rat_diff_cv}a. %
On the one hand, as we show in Fig.~\ref{fig:rat_diff_nn}a, the peak temperature $T^*$ gets increasingly closer to $T=1$ for \textit{any} value of $\langle CV\rangle$ as the number of neurons increases (while the point with $T^*\simeq 1$ remains relatively $N$-independent). %
On the other hand, this increasing proximity between $T^*$ and $T=1$ should be interpreted with caution.
Whether $T^*-1$ is small or large depends on a comparison with some natural scale of temperature variation in the problem.

We propose to compare $T^*-1$ with the width $\Delta T$ of the bell-shaped $c(T)$ curve at half height, 
\begin{equation}
\label{eq:DeltaT}
    \Delta T \equiv T^*-T_{1/2}\; ,
\end{equation}
where $T_{1/2}$ is defined by $c(T_{1/2})=c(T^*)/2$, as shown in Fig.~\ref{fig:rat_diff_cv}a.
Fig.~\ref{fig:rat_diff_nn}b shows that the $c(T)$ curves become sharper as $N$ increases, thus providing a natural scale with which to compare the results in Fig.~\ref{fig:rat_diff_nn}a. 
We thus define a \textit{normalized distance to criticality} $\tau$, defined as
\begin{equation}
\tau \equiv \frac{T^*-1}{\Delta T} \; . \label{eq:NDC}
\end{equation}

\begin{figure}[!ht]
    \centering
    \includegraphics[width=0.95\linewidth]{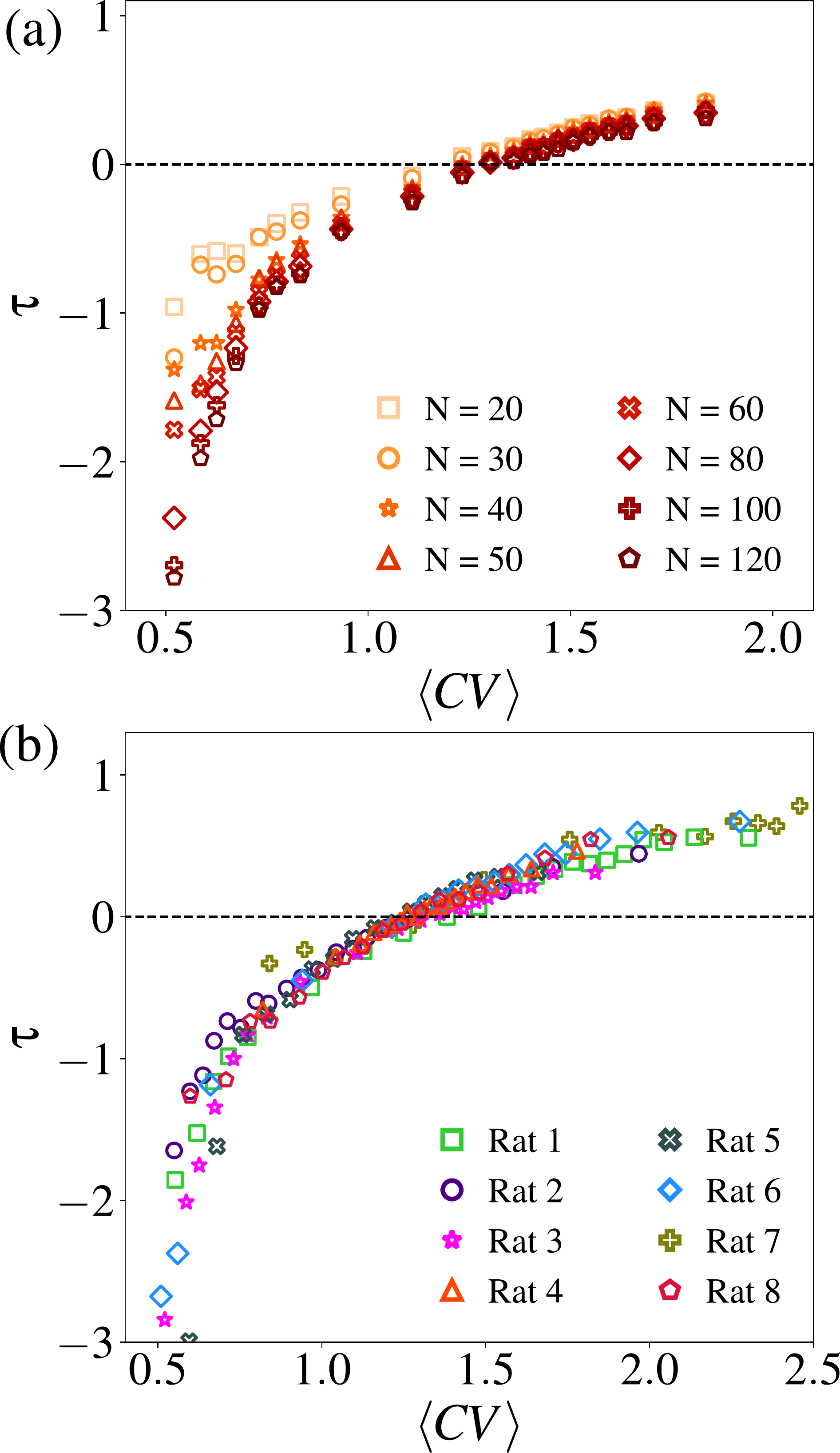}
    \caption{(a) Normalized distance to criticality $\tau$ versus $\langle CV\rangle$ quickly converges to a well-behaved function for increasing $N$ (Rat 3, $N=130$, $\Delta t=50$~ms). 
    (b) $\tau$ versus $\langle CV\rangle$ for the maximum number of neurons in each data set. $\tau$ crosses zero at approximately the same value of $\langle CV\rangle$ for all rats (see text for details). }
    \label{fig:rat_diff_cv1}
\end{figure}

Differently from the behavior of $T^*-1$ (Fig.~\ref{fig:rat_diff_nn}), $\tau$ as a function of $\langle CV\rangle$ converges quickly to a well-behaved function even for a fraction of the total number of neurons (Fig.~\ref{fig:rat_diff_cv1}a). 
Besides, this function $\tau(\langle CV\rangle)$ seems to be universal for this setup, in the sense that it is reproduced by different rats with different numbers of neurons (Fig.~\ref{fig:rat_diff_cv1}b). %
In particular, for all rats, $\tau$ crosses zero in approximately the same critical value of $\langle CV\rangle^* \approx 1.28 \pm 0.08$. 
This crossing is a strong indicator of universal behavior segregating regimes of low temperatures (synchronized states) and high temperature (desynchronized states).

%\section{Conclusion}

In conclusion, we have applied the Maximum Entropy approach of Mora \textit{et al.}, which takes into account the dynamical aspects of networks activity, to cortical spiking data of urethane-anesthetized rats. 
Since spiking variability undergoes major changes along the hours of the experiments, the data sets were parsed by $\langle CV \rangle$ in an attempt to fulfill, for each $\langle CV \rangle$, the stationarity required by the model.

When analyzed in this way, the method reveals signatures of criticality for a very narrow range of $\langle CV \rangle$ values. 
For very low (high) $\langle CV \rangle$, the system is more desynchronized (synchronized), which corresponds to a disordered (ordered) phase, i.e. with $T=1 > T^*$ ($T=1 < T^*$). 
We introduced a normalized distance to criticality $\tau$ whose behavior was universal across rats, consistently crossing zero at the same critical value $\langle CV\rangle^*$. 
These results are not reproduced by shuffled data and, as shown in Supplementary Material Section D~\cite{SM}, stand robust against changes in the time bin $\Delta t$ used to calculate firing rates (Fig.~S3~\cite{SM}), the order $v$ of the model (Fig.~S4~\cite{SM}) and the width $W$ of the windows employed to calculate $CV$ (Fig.~S5~\cite{SM}).

The critical value $\langle CV\rangle^*$ obtained with the Maximum Entropy approach is compatible (within error bars) with the one obtained by Fontenele \textit{et al.} ($\langle CV\rangle^* \approx 1.4 \pm 0.2 $) via neuronal avalanche scaling analysis~\cite{fontenele2019criticality}. 
Despite the completely different nature of these two approaches, both strongly suggest that a phase transition occurs at an intermediate level of synchronization for urethane-anesthetized rats.

%\section{Acknowledgment}
We acknowledge Thierry Mora and Olivier Marre for sharing the maximum entropy analysis code and fruitful discussion.
NL is thankful to FACEPE (Grant No. BCT-0426-1.05/18) and CAPES (Grant No. 88887.308754/2018-00) for their support. 
MC and PVC acknowledge support from CAPES (PROEX 534/2018 Grant No. 23038.003382/2018-39), FACEPE (Grant No. APQ-0642-1.05/18), and CNPq (Grants No. 301744/2018-1 and No. 425329/2018-6). This article was produced as part of the activities of FAPESP Center for Neuromathematics (Grant No. 2013/07699-0, S. Paulo Research Foundation). C.S.-C. and B.C. acknowledge support from FCT (Grants No. SFRH/BD/51992/2012 and No. SFRH/BD/98675/2013) and PAC, MEDPERSYST Project POCI-01-0145-FEDER-016428 (Portugal 2020).
A.J.R. received support from an FCT Investigator Fellow (IF/00883/2013) and acknowledges the Janssen Neuroscience Prize (first edition) and the BIAL Grant No. 30/2016. 
%\textcolor{red}{[OTHER GRANTS HERE!]}

\bibliographystyle{ieeetr}

\bibliography{name1}

\end{document}

\title{Supplemental Material for\\
``Signatures of brain criticality unveiled by maximum entropy analysis across cortical states''}  

\author{Nastaran Lotfi}
\email{nastaran@df.ufpe.br}
\affiliation{Departamento de Física, Universidade Federal de Pernambuco, Recife, PE 50670-901, Brazil}
\author{Antonio J. Fontenele}
\author{Tha{\'\i}s Feliciano}
\author{Leandro A. A. Aguiar}
\affiliation{Departamento de Física, Universidade Federal de Pernambuco, Recife, PE 50670-901, Brazil}
\author{Nivaldo A. P. de Vasconcelos}
\author{Carina Soares-Cunha}
\author{B\'arbara Coimbra}
\author{Ana Jo\~ao Rodrigues}
\author{Nuno Sousa}
\affiliation{Life and Health Sciences Research Institute (ICVS), School of Medicine, University of Minho, Braga, 4710-057, Portugal}
\affiliation{ICVS/3B’s - PT Government Associate Laboratory, Braga/Guimar\~aes, Portugal}
\author{Mauro Copelli}
\affiliation{Departamento de Física, Universidade Federal de Pernambuco, Recife, PE 50670-901, Brazil}
\author{Pedro V. Carelli}
\email{pedro.carelli@ufpe.br}
\affiliation{Departamento de Física, Universidade Federal de Pernambuco, Recife, PE 50670-901, Brazil}

\date{\today}
\maketitle
\newpage

\subsection{Data Acquisition}\label{sec:data-acquisition}

The data used in this analysis are taken from two experimental setups. 
As it has been described previously~\cite{fontenele2019criticality}, three Long-Evans rats, male, 250-360 g, 3-4 months old (five Wistar-Han rats, male, 350-500 g, 3-6 months old, Charles River) were used in the recordings. 
Animals were anaesthetized with 1.58 g/kg (1.44 g/kg) of fresh urethane, diluted at 20 $\%$ in saline, in 3 injections (i.p.), 15 min apart.

We implanted 32-(64-)channel silicon probes (BuzsakiA32/BuzsakiA64sp, Neuronexus), which are composed by 4 (6) shanks with 8 (10) sites/shank with impedance of 1-3~MOhm at 1~kHz, in the primary visual cortex of the rats (V1, Bregma: AP = -7.2, ML = 3.5). 
Shanks were 200 $\mu \text{m}$ apart and the area of each site was 160 $\mu\text{m}^2$, disposed from the tip in a staggered configuration, 20~$\mu \text{m}$ apart. 
All data were sampled at 24~(30)~kHz, amplified and digitized in a PZ2 TDT, which transmits to a RZ2 TDT base station (amplified and digitized in a single head-stage Intan RHD2164). 
All recordings were analyzed up to a duration of 3~hours. 

After recordings, spike sorting was performed by using the Klusta-Team software~\cite{kadir2014high,rossant2016spike} on raw electrophysiological data. 
Housing, surgical and recording procedures were in strict accordance with the CONCEA - MCTI, and was approved by the Federal University of Pernambuco (UFPE) Committee for Ethics in Animal Experimentation (23076.030111/2013-95 and 12/2015) and European Regulations (European Union Directive 2010/63/EU)).

\subsection{Surrogate Data}\label{sec:surrogate-data}
\begin{figure}[!htb]
    \centering
    \includegraphics[width=0.95\linewidth]{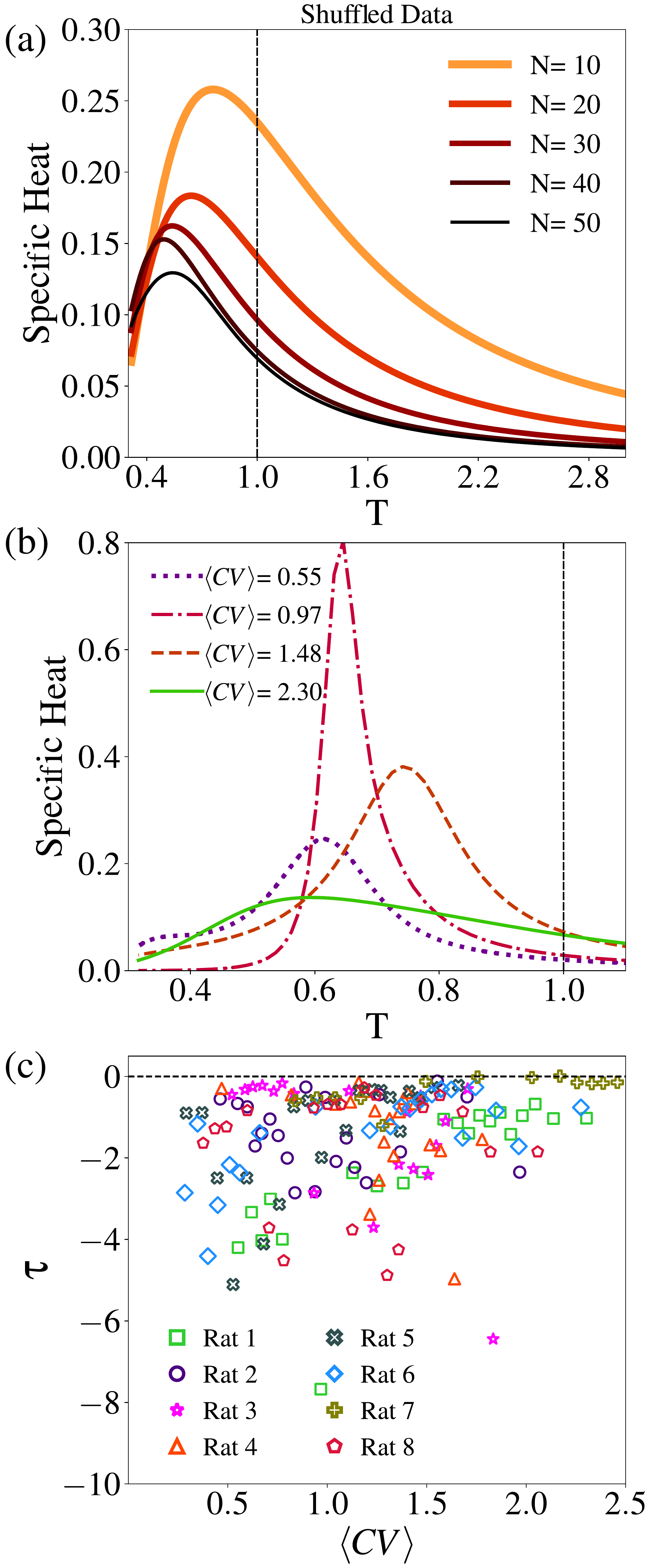}
    \caption{Variation of specific heat for shuffled data for Rat 1 ($\Delta t=50$~ms). (a) Results for the whole time series with different number of neurons. In (b) We illustrate the influence of $\langle CV\rangle$ on specific heat for a few examples with the maximum number of neurons. In (c) we show a scatter plot of normalized distance to criticality $\tau$ versus $\langle CV\rangle$ for all rats, breaking the universal structure shown in Fig.\ref{fig:rat_diff_N-cv}b. }
    \label{fig:diff_n_cv_shuffle}
\end{figure}
To test the significance of our results presented in the main text, we repeated the analysis over surrogate data. To obtain these series each neuron spike timing was randomized, keeping the total number of events fixed.

In the whole time series of the shuffled data sets, we verify that the specific heat remains with small values, its peak reduce when the number of neurons increase (in opposition to specific heat increase with original spiking series in Fig.~1a), and specific heat peak is typically very far from T=1 (Fig. \ref{fig:diff_n_cv_shuffle}a).

Repeating our maximum entropy analysis as a function of the cortical state, $\langle CV\rangle$, for surrogate data we verify that there is no specific trend in the behaviour of the specific heat Fig.~\ref{fig:diff_n_cv_shuffle}b (in comparison to Fig.~3a). To emphasize the inconsistency in behaviour of the shuffled data, we test for all rats and find $\tau$, Fig.~\ref{fig:diff_n_cv_shuffle}c. This plot shows that the universal trend in Fig.~3b disappears when the data is randomized.

\subsection{ \textit{CV} time series}\label{cv-variation}

\begin{figure}[!ht]
    \includegraphics[width=0.95\linewidth]{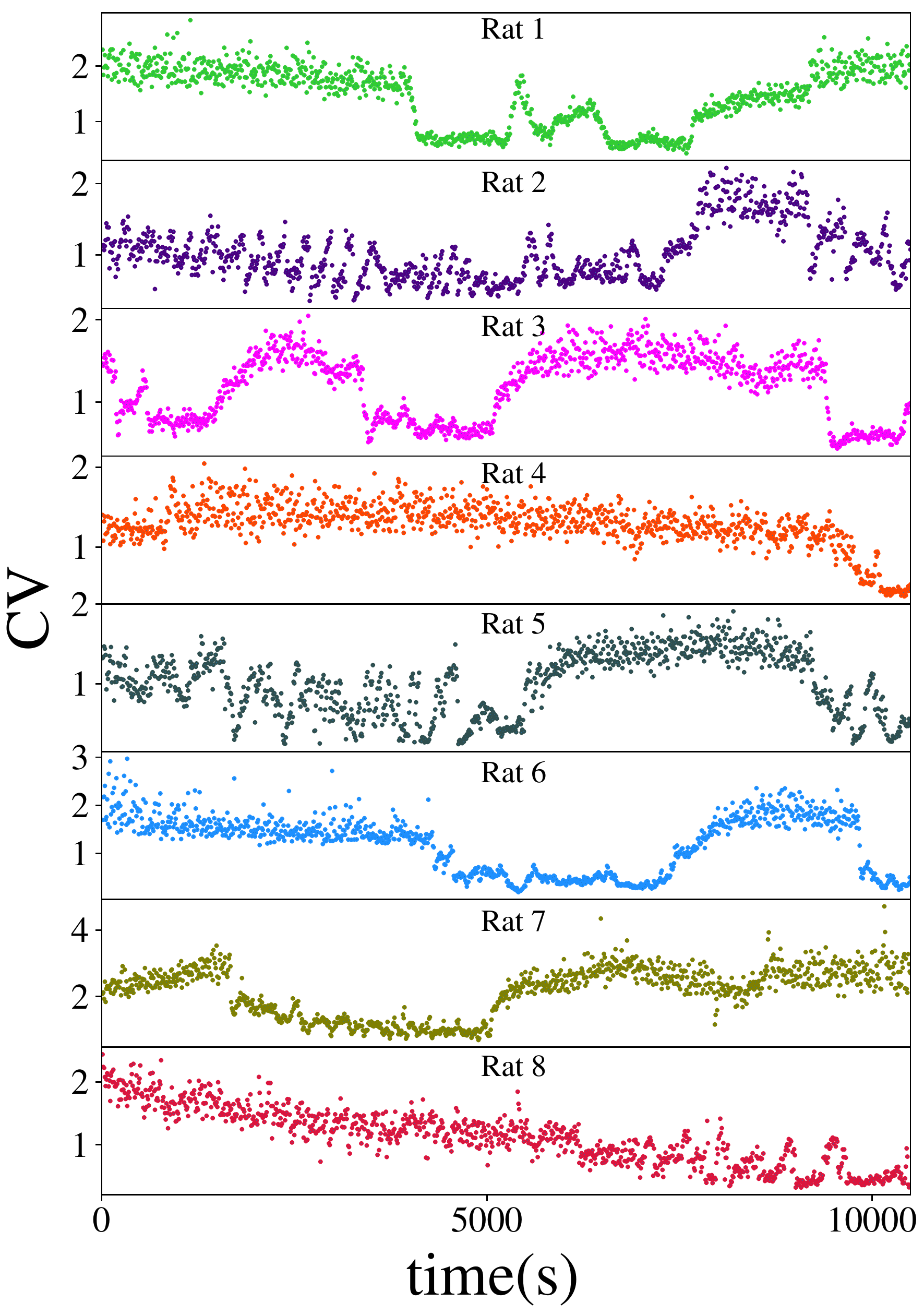}  
    \caption{Time series of \textit{CV} for different recordings ($\Delta t=50$~ms). 
    }
    \label{fig:cv_8}
\end{figure}

In Fig.~\ref{fig:cv_8}, we show time series of the \textit{CV} during 3~hours of continuous recording for different rats. It can be seen that in all the rats, there are variations from low to high \textit{CV}. This suggests in segmenting the data due to the firing rates for studying the maximum entropy.

\subsection{Robustness of the results}\label{deltat}

To further probe the robustness of our results, we also repeated our analysis with different time scales for discretization of the neural firing and parsing the cortical states.

\subsubsection{Time resolution for the firing rates}

\begin{figure}[!ht]
    \centering

    \includegraphics[width=0.95\linewidth]{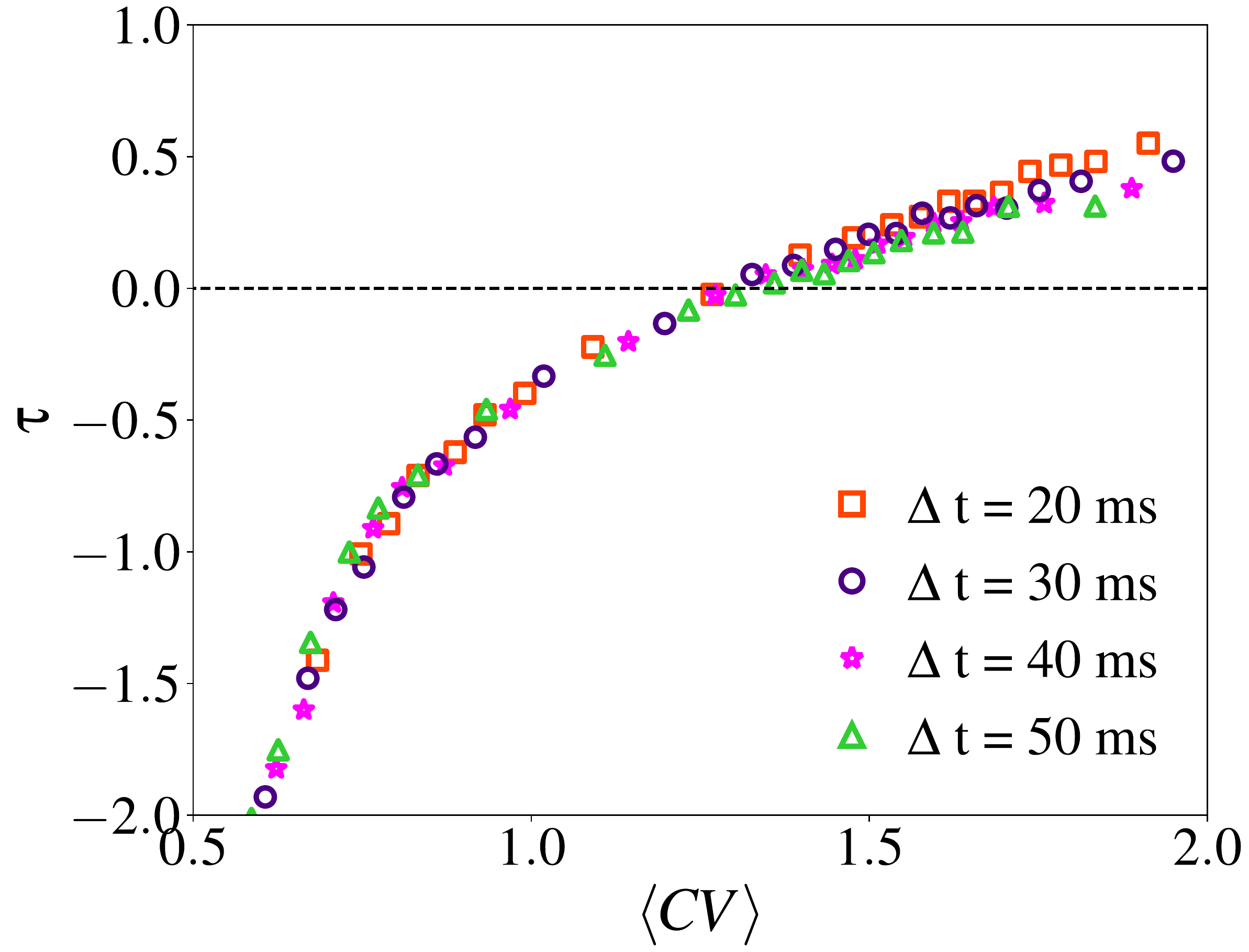}
    \caption {Robustness of the behavior of normalized distance to criticality ($\tau$) versus $\langle CV\rangle$, considering different time bins ($\Delta t$) for calculating firing rates.}
    \label{fig:rat_diff_ddt-cv}
\end{figure}

The first temporal parameter we have modified was the resolution, $\Delta t$, used in the calculation of the firing rates. In Fig.~\ref{fig:rat_diff_ddt-cv}, different values are selected for $\Delta t$, and maximum entropy analysis is done with variation of $\langle CV\rangle$. It can be seen that all curves are collapsing together and the results are robust against these modification.

\begin{figure}[!ht]
    \centering
    \includegraphics[width=0.95\linewidth]{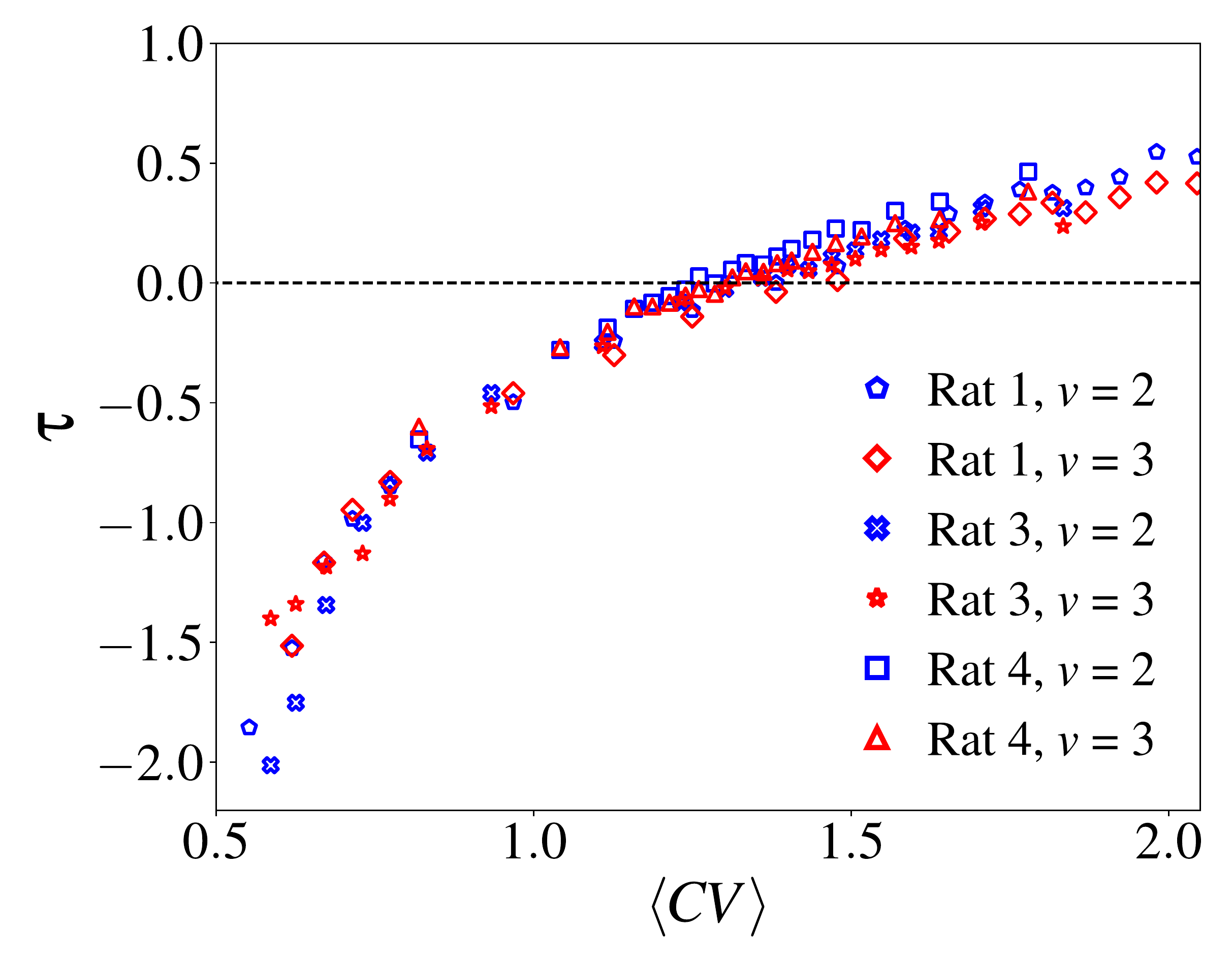}

    \caption{Robustness of the behaviour of the normalized distance to criticality ($\tau$) as a function of $\langle CV\rangle$ for different values of the model order ($v$). We show results from three typical rats using $v=2$ and $v=3$ ($\Delta t=50$~ms).}
    \label{fig:diff_u}
\end{figure}
\begin{figure}[!htb]
    \centering
    \includegraphics[width=0.95\linewidth]{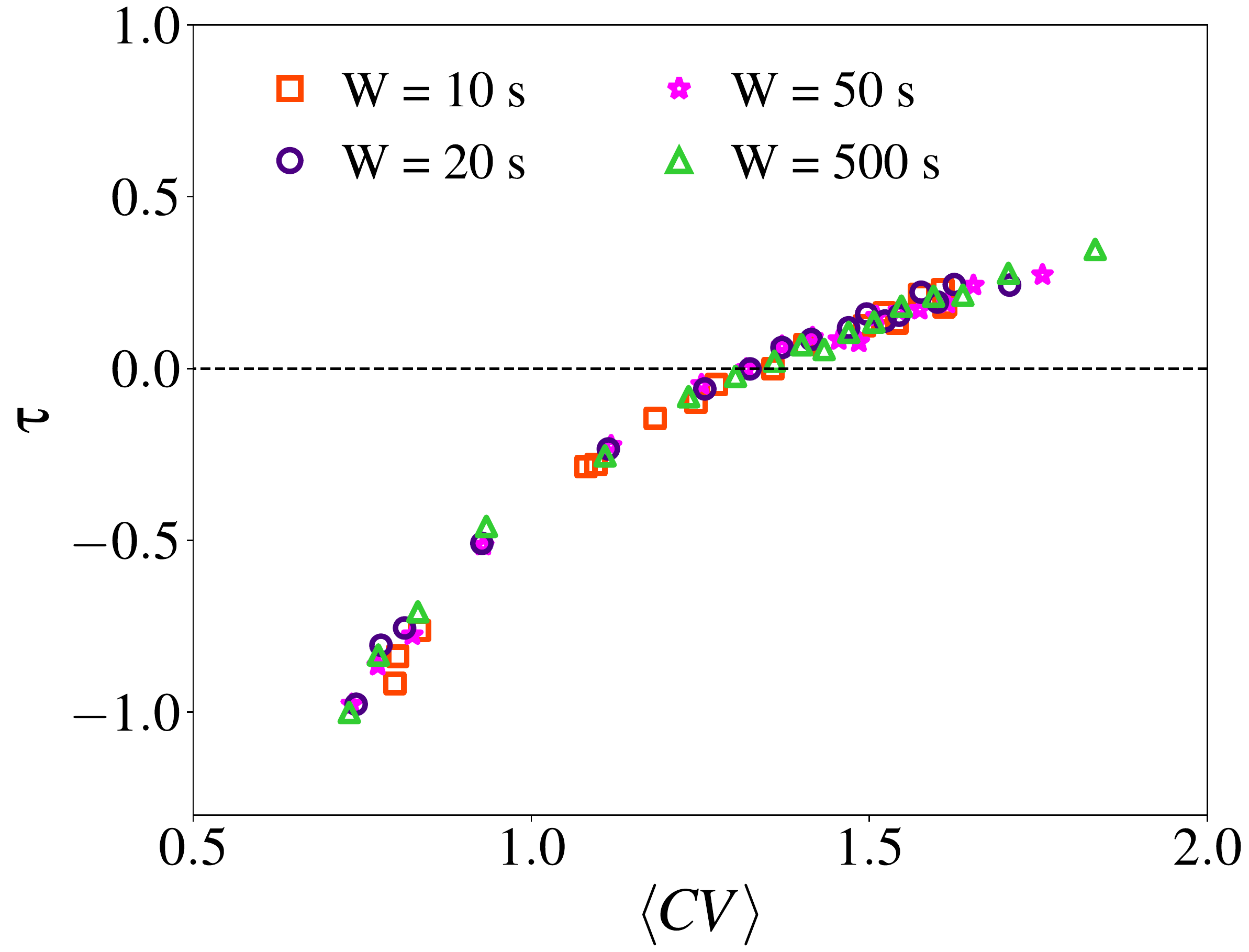}
    \caption {Universal behavior of normalized distance to criticality ($\tau$) versus $\langle CV\rangle$ for different time scales used for calculating \textit{CV}.}
    \label{fig:rat_diff_N-cv}
\end{figure}

\subsubsection{Model order \textit{v}}\label{changing-cv}

Another parameter of the model is its temporal order $v$ that defines how many time steps are considered for the model dynamics (see eq.~(2)). 
We show that changing this value does not affect the results, which lead to similar trends, Fig.~\ref{fig:diff_u} (results presented for three randomly selected rats).

\subsubsection{Time resolution for defining a cortical state}\label{time-window}

We have also explored different values for the time scale used to define a cortical state. For calculating \textit{CV}, a time windows of $W=10$ s was used during the mains text results. Changing this value to the larger windows, the obtained results for maximum entropy analysis are showing very similar behaviour Fig.~\ref{fig:rat_diff_N-cv}. 

\bibliographystyle{ieeetr}
\bibliography{name1}